# Swift heavy ions in polyethylene: simulation of damage formation along the path

P. Babaev[1,*], R. Voronkov[1], A.E. Volkov[1]

[1]*P.N. Lebedev Physical Institute of the Russian Academy of Sciences, Leninskij pr., 53,119991 Moscow, Russia*

[*]*babaevpa@lebedev.ru*

## Abstract

We present results of atomic-level simulations of damage formation along the paths of swift heavy ions (SHI) decelerated in the electronic stopping regime in amorphous polyethylene. The applied model combines the Monte-Carlo code TREKIS-3, which describes excitation of the electronic and atomic systems around the ion trajectory, with molecular dynamics simulations of the response of the atomic system to the excitation. The simulation results were used to reconstruct the damage configuration, shape and size of the damaged region. We demonstrated that the positions of the maximum energy loss and maximum damage on the ion trajectory do not coincide, being separated by more than 10 micrometers. The difference between the thresholds of damage production by ions with energies realizing the opposite shoulders of the Bragg curve of the electronic stopping was found. We also analyzed the spatial distribution of chemically active fragments of polyethylene chains formed around the ion trajectory as a function of SHI energy.

## I. Introduction

Releasing an extreme energy into the electronic system of a solid, swift heavy ions (SHI, $E > 1$ MeV/a.m.u., $M \geq 10$ a.m.u.) can form straight damaged tracks inside the material with the high length-to-radius ratio (~$10^4$-$10^5$). It makes SHI irradiation a unique tool for modifications of bulk targets. Atomic-scale simulations [1] are effective for understanding the mechanisms of SHI track formation, boosting development of techniques for nanostructuring, membrane creation, cancer therapy, particle detection, etc. [1,2]. Despite the importance of polymers for applications, attempts at quantitative all-atom modeling of the SHI track formation in these materials have appeared only recently [3,4]. This is explained by a lack of quantitative models of material excitation around the ion path as well as difficulties in development of force fields describing the atomic dynamics in organic solids under extreme conditions. In order to bypass at least the second problem, the authors of [5,6] used a coarse-grained approach in their simulations of polymers. Considering the polymer structure as a long chain of monomer beads, they demonstrated an ability to simulate large structural changes such as craters and rims on the surface of the irradiated sample. All-atom and coarse-grained approaches were also applied in simulations of polymers under atomic cluster irradiation and in sputtering modeling [7,8].

Very often, the two-temperature thermal spike model (TTM) [3,9] is used to describe the excitation of materials around the ion trajectory, providing the initial conditions for MD modeling. Unfortunately, the TTM cannot provide realistic radial distributions of the energy density deposited into the electronic and ionic systems of the target, consistent with the energy of an incident SHI [1]. In particular, it does not take into account the effect of the spectrum of electrons generated due to target ionization by an SHI. This is most clearly manifested in the position of the maximum damage on the ion trajectory which does not coincide with the position of the deposited energy maximum [10], as well as in different damages produced by the same ions of different energies but providing the same electron energy loss ($S_e$) values on the opposite shoulders of the Bragg peak of the energy dependence of the ion electronic stopping curve (the so-called "velocity effect") [11].

In this work, we study the dependence of damaging of amorphous polyethylene on the SHI velocity along the whole ion trajectory. We combine the Monte-Carlo (MC) code TREKIS-3 [12,13], which describes the excitation of the electronic and atomic systems of the target around the ion trajectory, with the molecular dynamics (MD) simulations (LAMMPS [14]), illustrating the atomic system response to the excitation. Within the Dynamical Structure Factor - Complex Dielectric Function (Loss Function) formalism [15], the cross sections used in TREKIS take into account the effects of the collective response of the atomic and electronic systems of a solid on the scattering of charged particles. The application of the reactive force field in MD simulations provides a description of both the breaking and the formation of chemical bonds during the evolution of the simulated atomic system of polyethylene around the SHI trajectory. The applied technique allowed us to determine the radial damage profiles along the entire ion trajectory, i.e. broken chemical bonds and chemically active fragments in each cross-sectional layer of the ion path.

We demonstrate that the positions of the structural and chemical damage maxima on the ion trajectory do not coincide with that of the deposited energy maximum, with spatial separations of at least 10 micrometers. We demonstrate the appearance of a loop-like dependence of the radius of the damaged region in amorphous polyethylene on the SHI electronic energy loss, with different thresholds for damage by the same ions with energies forming the left and right shoulders of the Bragg curve.

## II. Model

As will be shown below, an SHI impact causes small changes in the density of the amorphous polyethylene around the ion trajectory in the simulated samples. Taking this into account, we use the density of unsaturated carbons in the track to measure material damage in the presented paper. As before in ref. [4], we define an unsaturated carbon atom as an atom with dangling bonds (or/and with double-triple bonding) and thus having less than 4 neighbors within the interaction cut-off (2 Å for C-C and 1.8 Å for C-H interactions using the AIREBO-M interatomic potential [16]).

To compute the damage along the trajectory, we perform simulations in several cross sections (transverse layers) along the SHI trajectory. Each layer in our simulations is represented by a polyethylene MD supercell, which is assumed to be located in the depth of the material, that enables application of the periodic boundary conditions along the ion trajectory to the MD supercell in the simulations. The boundaries of the supercell in the transverse directions provide negligible relaxation effects in the material due to scattering of rare fast electrons or diverging weak elastic waves, so the periodic conditions are also established at these boundaries. On the other hand, such transverse dimensions allow to analyze the formation of even distant, unsaturated carbons. The computation time of each simulation was 250 ps, since further changes in the spatial distribution and the number of unsaturated carbons do not occur, as shown in [4].

Finally, the amorphous polyethylene supercell used in each simulation has dimensions of $204 \times 190 \times 50$ Å$^3$ and contains 207168 atoms. It is composed of polymer chains. They are made up of $CH_2$ monomers and have $CH_3$ groups at their ends. Each chain contains 442 carbons and 886 hydrogens for a total of 1328 atoms. The supercell was prepared with the EMC tool [40] and has the density of 0.82 g/cm$^3$ [17]. The AIREBO-M force field [16], which takes into account the breaking/formation of chemical bonds, is used to describe the many-particle interactions between atoms in PE.

The obtained spatial damage distributions in the all transverse layers are smoothly approximated between the layers restoring the shape and structure of the damaged region along the whole ion trajectory (ion track). This method of damaged track description has previously been successfully applied to inorganic dielectrics [1,10].

The event-by-event Monte Carlo code TREKIS-3 [8] describes the passage of the decelerating SHI and the subsequent excitations of the electronic and atomic systems up to 100 fs after the ion impact, until the excited electronic system cools down. Calculations with the TREKIS-3 code of the ion energy loss allow to obtain the energy of the heavy projectile in every point of its trajectory. This knowledge enables quantitative descriptions with TREKIS-3 of excitation of the electronic and atomic system of the target in each transverse layer. Within the first-order Born approximation, TREKIS-3 uses the scattering cross sections of charged particles expressed in terms of the energy loss function (ELF) of a target [12], which takes into account the collective response of the electronic systems of polyethylene to the excitation. We restore the ELF from the experimental optical data [15]. The Rutherford cross sections with the modified Molier screening parameter [12,18] describe the scattering of electrons and valence holes on target atoms [7,9] resulting in energy transfer to the atomic system of the material. The Monte Carlo procedure was iterated $10^3$ times to obtain reliable statistics.

The interatomic potential of a solid transiently changes due to extreme electronic excitation in the proximity of the ion trajectory [1]. To take into account the energy transferred into the atomic system due to the non-thermal acceleration of the target atoms caused by this change, we converted the potential energy accumulated in the generated electron-hole pairs into the kinetic energy of the atomic system at 100 fs after the SHI passage [19,20]. It has been shown many times that the summation of the energy received by the atoms via this channel with the energy from electron and valence hole scattering provides the realistic density of the kinetic energy received by the atomic system around the ion trajectory [18,21].

The density of the energy deposited into the atomic system by the time of the electronic system cooling down forms the initial condition for the molecular dynamics modeling with the LAMMPS code of the atomic system response to the excitation[14]. Due to the randomization of the transferred momenta, the initial velocities of the excited atoms in the coaxial cylindrical layers surrounding the ion trajectory are set at fixed deposited energy specific for the layer assuming a Gaussian-like dispersion of the kinetic energy and the uniform spatial distribution of the atomic momenta within each layer [22].

### III. Results

Tables 1 and 2 present the irradiation parameters of Xe ions in the energy range from 9.4 to 2318 MeV and U ions in the energy range from 14 to 19336 MeV, respectively.

**Table 1.** Irradiation parameters of Xe ions used in the simulations. The table displays the energy losses and ion ranges calculated by TREKIS-3, as well as the number of unsaturated carbons obtained from the MD analysis for each ion energy.

| Energy, MeV | $dE/dx$ TREKIS-3, keV/nm | Residual range TREKIS-3, μm | Number of unsaturated carbons |
|---|---|---|---|
| 9.4 | 3 | 0 | 0 |
| 30 | 5.94 | 4.5 | 2926 |
| 60 | 7.21 | 9.1 | 3543 |
| 100 | 8.24 | 14.3 | 3914 |
| 160 | 8.75 | 21.2 | 3940 |
| 188 | 9 | 24.4 | 3972 |
| 300 | 9.16 | 36.8 | 3740 |
| 885 | 7.92 | 105 | 2277 |
| 1470 | 6.9 | 189.3 | 1355 |
| 1888 | 6.22 | 251 | 879 |
| 2318 | 5.7 | 325.5 | 0 |

**Table 2.** Irradiation parameters of U ions used in the simulations. The table displays the energy losses and ion ranges calculated with TREKIS-3, as well as the number of unsaturated carbons obtained from the MD analysis for each ion energy.

| Energy, MeV | dE/dx TREKIS-3, keV/nm | Residual range TREKIS, μm | Number of unsaturated carbons |
|---|---|---|---|
| 14 | 3 | 0 | 0 |
| 30 | 6.9 | 4.2 | 4399 |
| 40 | 7.9 | 5.5 | 5524 |
| 50 | 8.7 | 6.7 | 6040 |
| 60 | 9.3 | 7.8 | 6414 |
| 100 | 10.9 | 11.8 | 7405 |
| 150 | 12.1 | 16 | 8185 |
| 205 | 13.4 | 20.5 | 8952 |
| 256 | 14 | 24.2 | 8986 |
| 299 | 14.5 | 27.3 | 9211 |
| 400 | 15.1 | 34.1 | 9100 |
| 500 | 15.4 | 40.7 | 9220 |
| 700 | 15.6 | 53.6 | 8756 |
| 1060 | 15.4 | 76.9 | 8201 |
| 1330 | 15.1 | 95.1 | 7555 |
| 2715 | 13.4 | 192.6 | 5939 |
| 4345 | 11.8 | 322.4 | 4539 |
| 5500 | 10.9 | 424.8 | 3889 |
| 7472 | 9.6 | 618.8 | 2906 |
| 9453 | 8.6 | 838.1 | 2186 |
| 14000 | 7.2 | 1506 | 1062 |
| 19336 | 5.7 | 2311.5 | 0 |

### III.1. Structure of tracks

Our calculations demonstrated that the material density in the damaged area does not change much and cannot be used as a criterion for the damage production. Therefore, we monitor unsaturated carbon atoms as an indicator of structural damage and chemically active polymer chains. Figures 1a and 1b demonstrate that for all energies, the damage caused by the ions accumulates in circular regions of approximately the same size, which does not exceed 20 nm and is consistent with the experimentally measured damage diameters in various polymers [23,24]. However, the volume density of the damage varies with the ion energy or different positions on the ion trajectory and decreases from the center to the periphery of the track. We observed, that unsaturated carbon atoms form small clusters rather than a solid track along the trajectory. Uranium ions produce more dense damage than that caused by xenons.

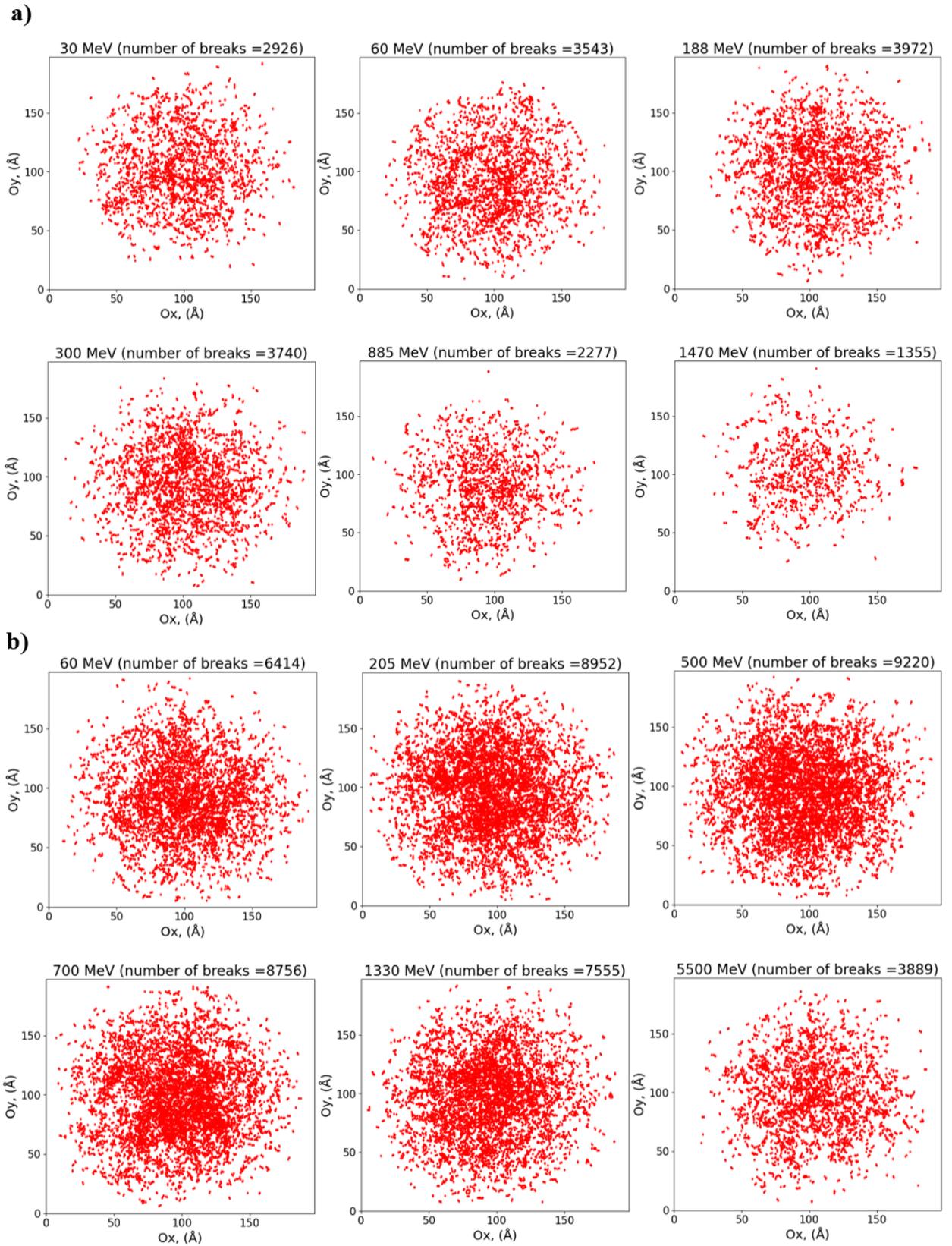

**Figure 1.** Unsaturated carbons after impact of **a)** Xe and **b)** U ions with different energies.

Figure 2 shows the sequence of the damaged transverse layers (for the simulated ion energies) along the trajectories of xenon and uranium ions. The damage profiles clearly demonstrate the structure of the damage at each point of the trajectory: the core containing highly damaged material (unsaturated carbons) surrounded by a halo of damaged clusters at the track periphery.

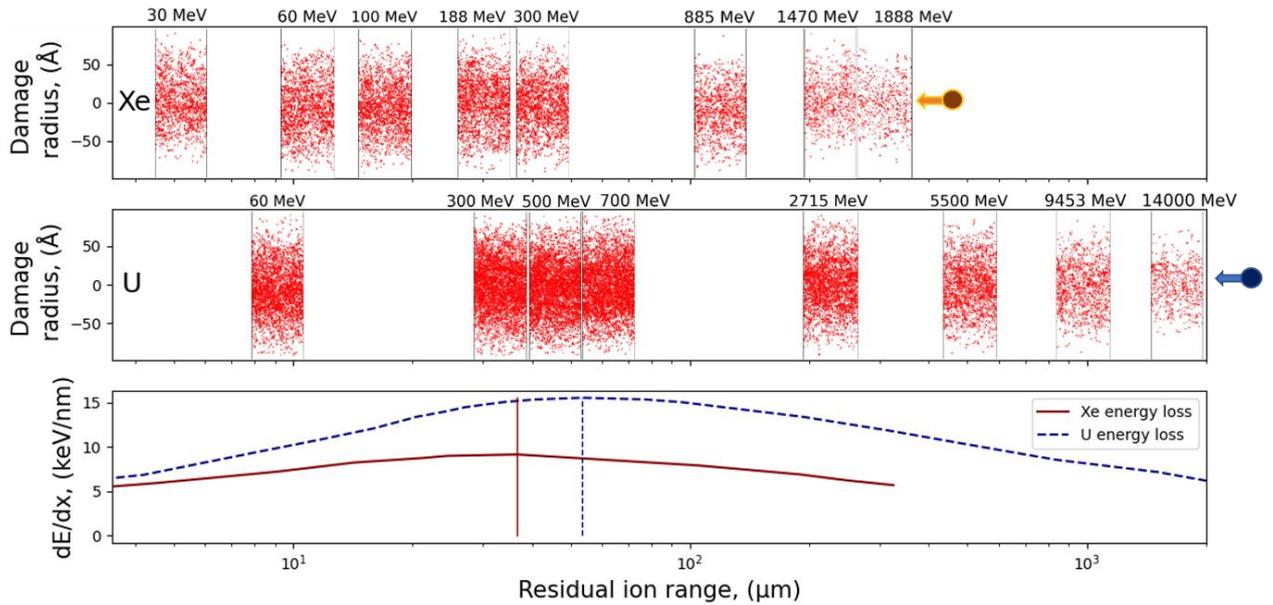

**Figure 2.** Damaged amorphous polyethylene along the 1888 MeV Xe and 1400 MeV U trajectories. The residual ion range (the distance before the ion stop point) is shown on a logarithmic scale. The width of each layer is 50 Å (layers are not to scale). The left edge of each slice is located at the point of the trajectory to which it corresponds. The energy loss of both ions as a function of the residual ion range is shown in the lower graph. The vertical lines at the bottom indicate the maximum energy loss for both ions.

We use the local density of unsaturated carbons in the coaxial cylindrical layers of the track to find the radius at which it falls below the threshold for each simulated supercell. To determine the density threshold value in the radial coaxial layers separating the track core from its halo, we rely on the experimental reference: 1470 MeV Xe ion forms a track in polyethylene with an average diameter of 60 Å. Comparing our simulation in a supercell with this ion energy gives a value of the volumetric threshold density of unsaturated carbons equal to 0.0027 1/Å$^3$ [25]. We define a halo as an area of lower damage density that is attached to the core but does not contain single unsaturated carbon atoms.

Figure 3 shows the core and halo radii along the trajectories of Xe and U ions. The incoming ions are slowed down, leaving an ever-increasing trail of damage. The lengths of the damaged tracks are about 250 μm for Xe and 2000 μm for U. The damaged track is asymmetric along the ion path with respect to the position of the Bragg peak, following the energy dependence of the electronic energy loss.

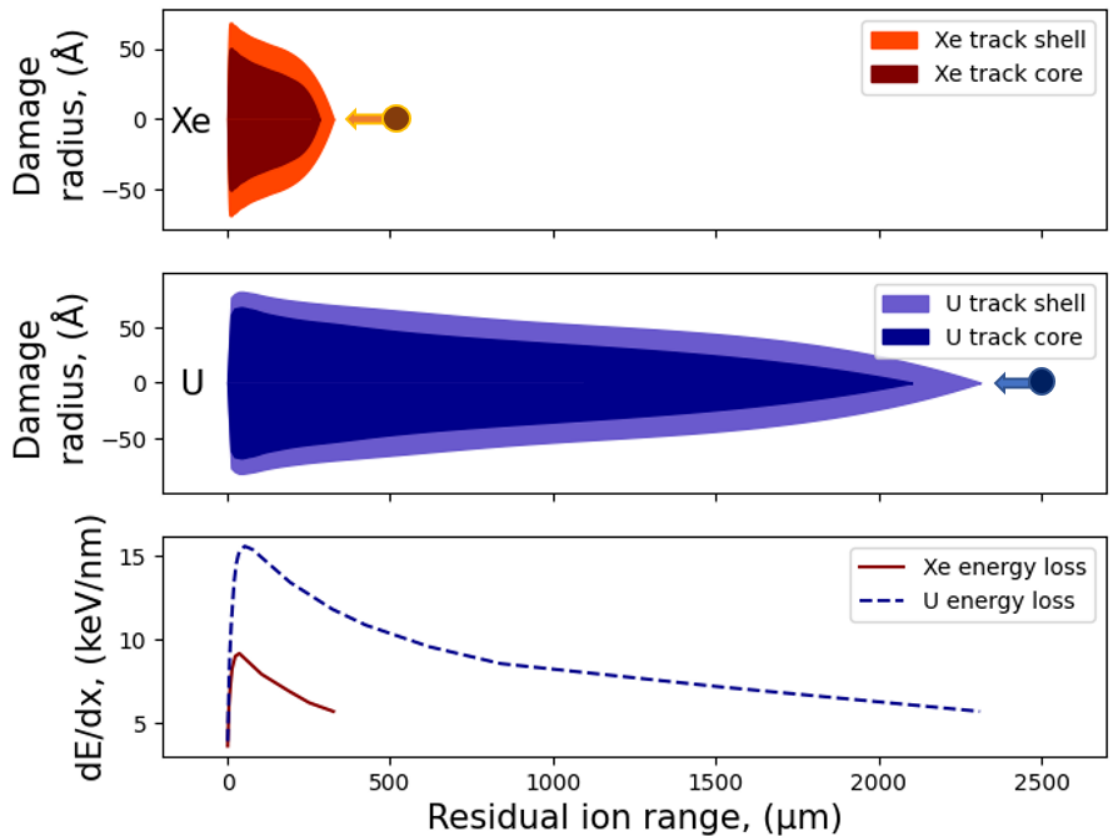

**Figure 3.** The radial distribution of the damage as a function of the residual range of Xe and U ions in polyethylene (upper panels). Arrows indicate the direction of SHI passage. The Bragg curves are shown for comparison (bottom panel).

### III.2. The velocity effect

Figure 4a demonstrates that the position of the damage maximum on the Xe trajectory does not coincide with that of the deposited energy. The maximum damage is separated by about 10 micrometers from the location of the deposited energy maximum. The same mismatch is observed when combining the dependences of the damage and the Xe energy loss on the ion energy (see Fig. 4b). The maximum damage corresponds to ions with an energy of 120 MeV, which is more than two times lower than the energy realizing the Bragg peak (300 MeV).

This difference reflects the so-called general "velocity effect", which means that the same ions with the same energy loss but different velocities cause different damage [10,11,26]. This effect is caused by different spectra of electrons excited by such ions: the higher the SHI velocity, the higher the maximum energy transferred to an electron. Faster electrons produced by the faster ion escape farther from the track, reducing the deposited energy density and the amount of damage in the center of the track.

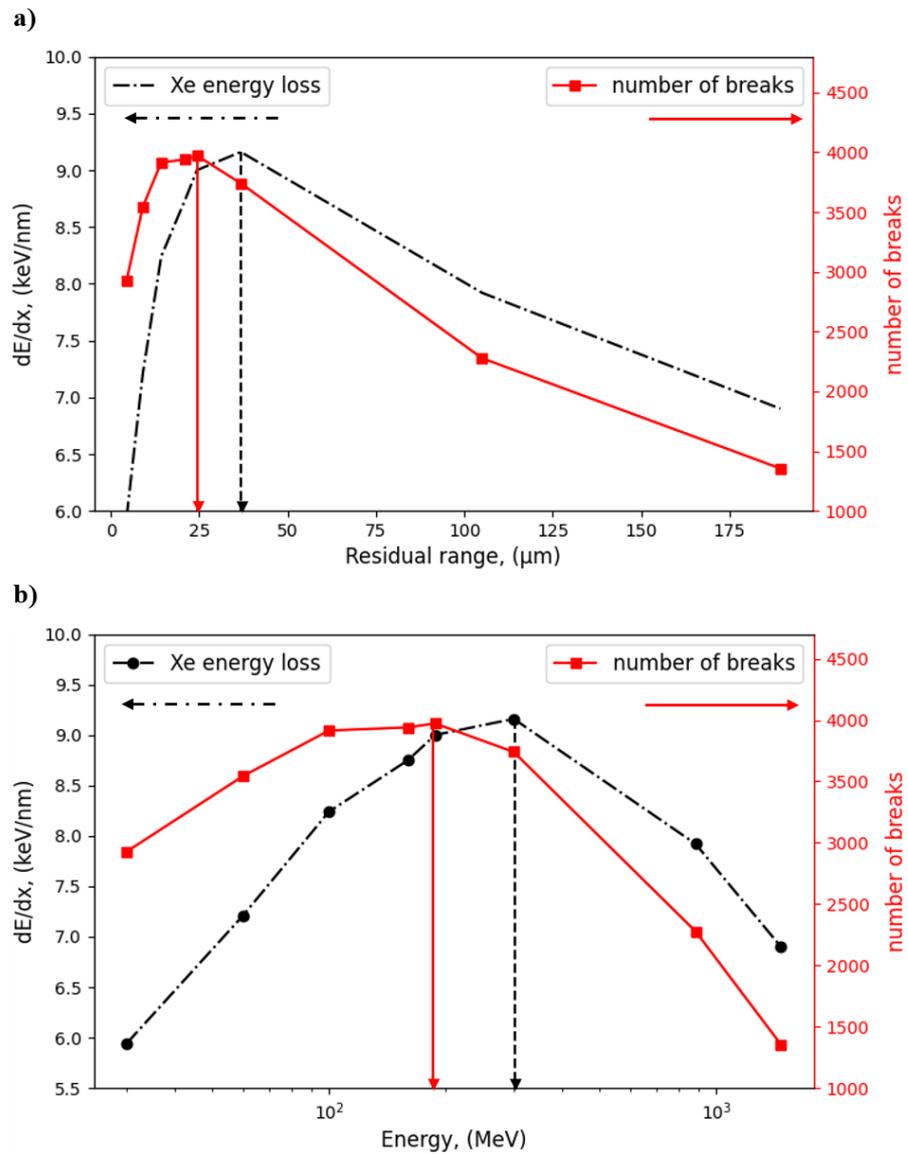

**Figure 4.** The dependence of the energy loss and the number of unsaturated carbons in the xenon track on the (a) residual range and the (b) SHI energy.

As in the case of the Xe ions, the position of the maximum damage on the uranium trajectory does not coincide with that of the U energy loss peak. The maximum damage is separated by about 13 μm from the location of the deposited energy maximum (see Fig. 5a). A similar difference is observed when comparing the dependences of the damage production and the energy loss on the ion energy. The maximum damage is caused by U ions with the energy lower than that of the Bragg peak (700 MeV) by 200 MeV, see Fig. 5b. Comparing the irradiation with Xe and U ions, the discrepancy increases with increasing of the SHI mass, in agreement with the previous studies in inorganic insulators [10,26]. Although it was previously reported in experimental studies that the velocity effect was not observed in polymeric materials [27,28], a recent detailed SAXS study confirms the presence of the effect in the polymer polyethylene terephthalate (PET) [23]. There is no doubt that the velocity effect can be detected in other polymers using advanced techniques.

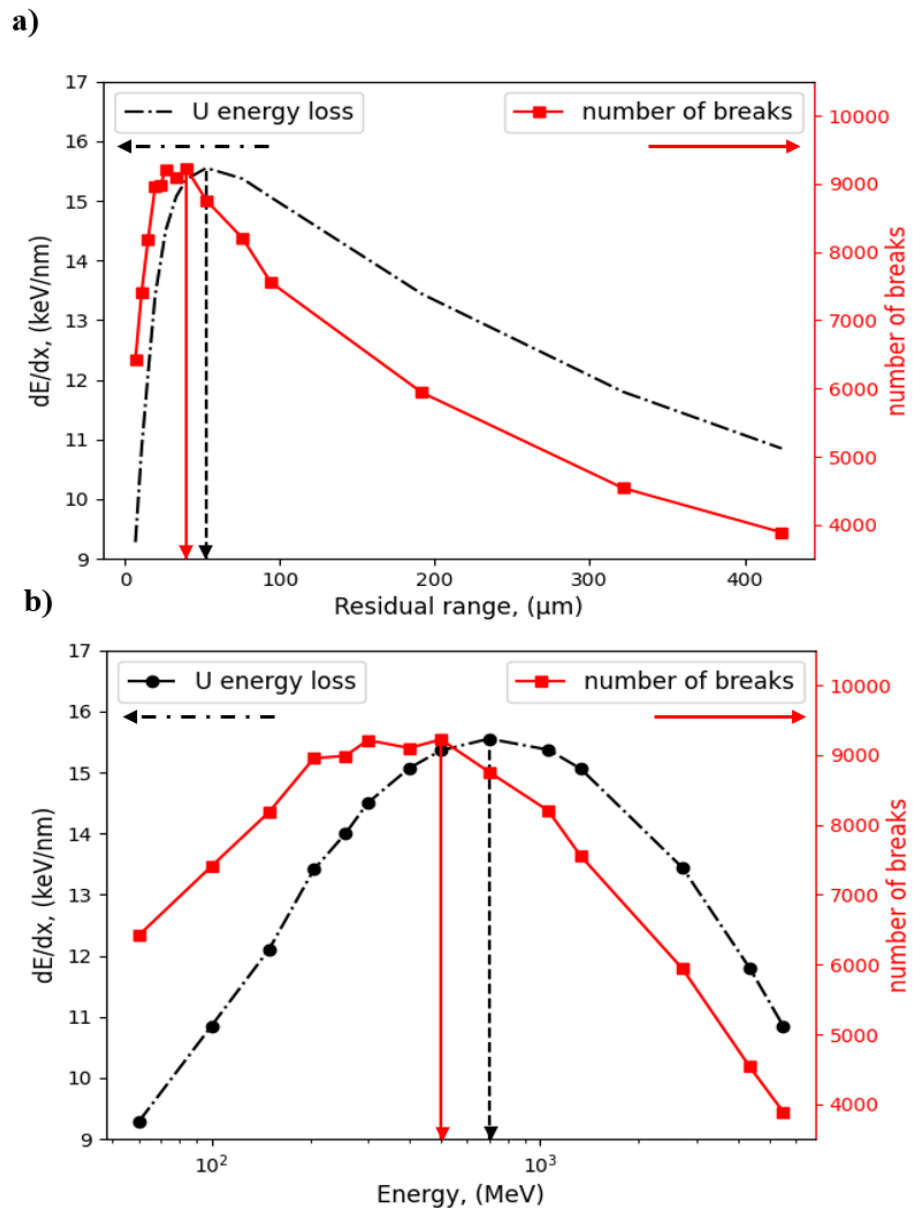

**Figure 5.** The dependence of the energy loss and the number of unsaturated carbons in the uranium track on the a) residual range and the b) SHI energy.

Figure 6 presents the loop-like dependence of the number of unsaturated carbons on the energy loss of Xe and U ions. The figure clearly shows that the amount of damage depends not only on the energy loss and velocity of the passing ion, but also on its mass, which is consistent with the previous results for dielectrics [10].

The graph can be categorized into two branches based on the energy of the SHI: "slow" and " fast " ions with the energy loss before and after the Bragg peak value, respectively. We approximate the dependencies with smooth curves to determine the threshold energy losses. Parabolas are used for the left branch: $y^2 = -47x^2 + 1607x - 4410$ (Uranium), $y^2 = -101x^2 + 1872x - 4661$ (Xenon). Straight lines approximate the right branch: $y = 753x - 4342$ (Uranium), $y = 1060x - 6013$ (Xenon). The blurred part of the curves corresponds to the ion energies at which the losses in the electronic subsystem cease to dominate over the nuclear losses. This part goes beyond the TREKIS-3 model and is extrapolated from the data for faster ions.

The curve fitting shows that the energy loss thresholds for damage production are approximately the same for Xe and U: 3 and 5.7 keV/nm for the lower and upper thresholds, respectively.

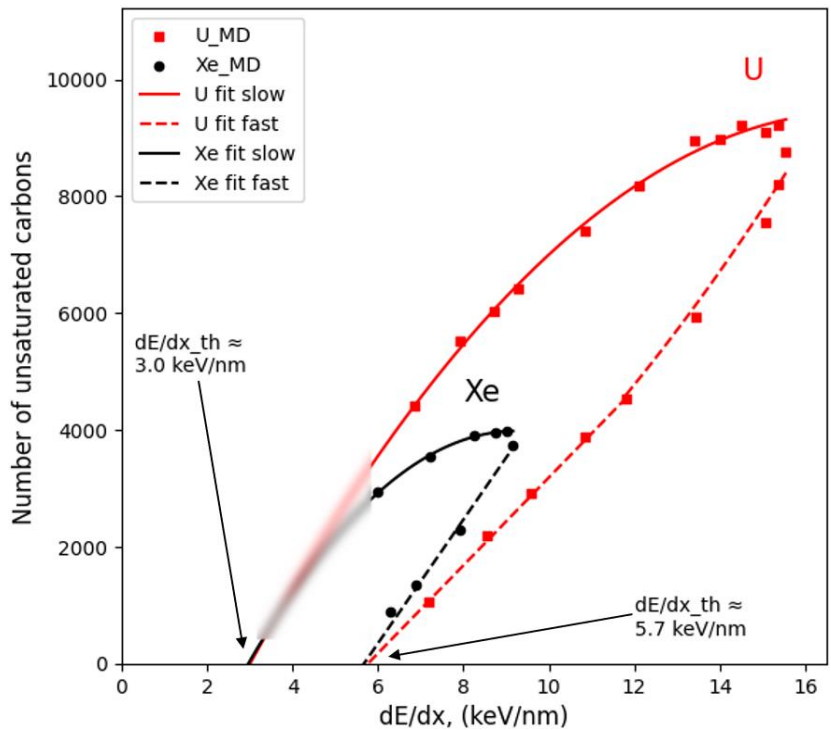

**Figure 6.** Relationship between the number of unsaturated carbons and the energy loss of uranium and xenon ions in amorphous polyethylene. The blurred part of the graph corresponds to the ion energies at which the nuclear energy losses become comparable to the electronic energy losses.

## III.3. Fragmentation of polyethylene chains

Rearrangement of atoms and bonds not only causes structural damage, but also changes the chemical composition and properties of the polyethylene target [2,29,30]. As shown in [4], almost all hydrocarbon species with masses less than that of the full chain are represented in the track after passing of an SHI.

Figure 9 shows the dependence of the number of the most abundant species in the relaxed uranium track on the ion energy. The peak profiles of the low mass fragments are qualitatively the same as the number of broken bonds in Figure 5 with the maximum around 300-400 MeV ion energy. However, the minimum for undamaged chains is located at 256 MeV, which is different from the positions of the maximum number of broken bonds and the Bragg peak.

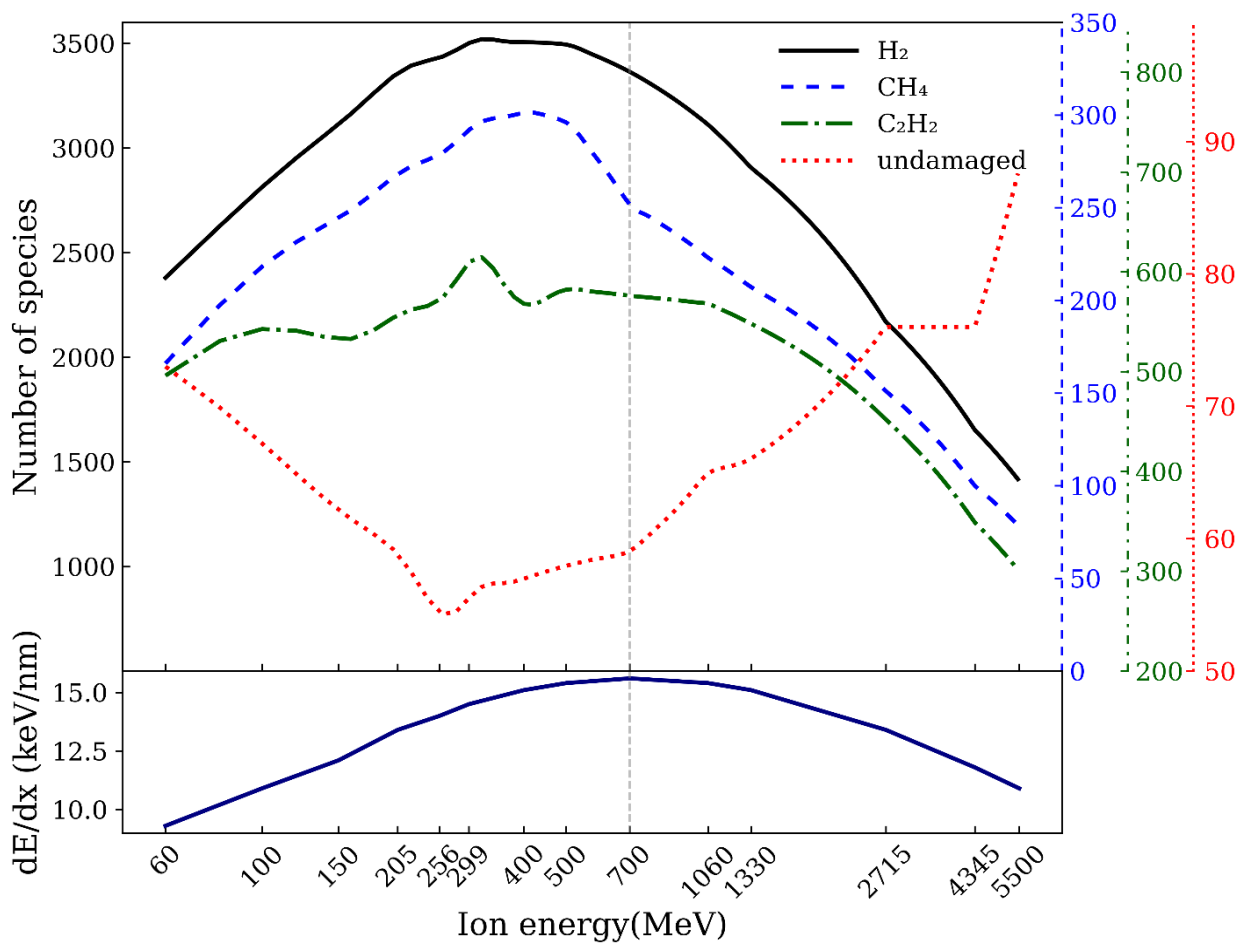

**Figure 7.** Dependence of the number of most redundant fragments (mass spectra) on U ion energy. The lower graph presents the U ion energy losses. The vertical line indicates the Bragg peak.

In addition to various hydrocarbon species, hydrocarbon free radicals (atoms with unpaired electrons) are formed. Species containing such radicals are extremely chemically active. Using the MD output with atomic coordinates, we identified such atoms using the following algorithm. In each fragment, we identified a sequence of carbon atoms with a reduced number of hydrogens. Then we calculated the number of missing hydrogens in the sequence. If the number of missing hydrogens in the sequence is even, we assume that corresponding "unpaired" electrons form double and triple bonds within the sequence and the whole sequence has no active atom. If the number is odd, we assume that there is an active atom randomly chosen within the sequence.

This algorithm can only be used for estimation, since it only works for simple hydrocarbon chains. The development of a precise algorithm taking into account all possible cases requires a separate dedicated study, which is beyond the scope of this paper. However, the data obtained in this study contain all the necessary information for the further work.

Keeping this in mind, we calculated the radial distributions of chemically active carbon atoms for each incident ion energy from Table 2 with 10 Å step (see Figure 8). It can be seen that such atoms are mainly concentrated in the region of ~40-60 Å near the ion trajectory with the maximum energy of the incident uranium ion of 1060 MeV, which does not coincide with either the Bragg peak (700 MeV) or the positions of maximum damage (500 MeV). A high level of excitation in the center of the track allows chemically active carbons to be neutralized, while excitation at the track periphery is only sufficient for the appearance of chemically active species.

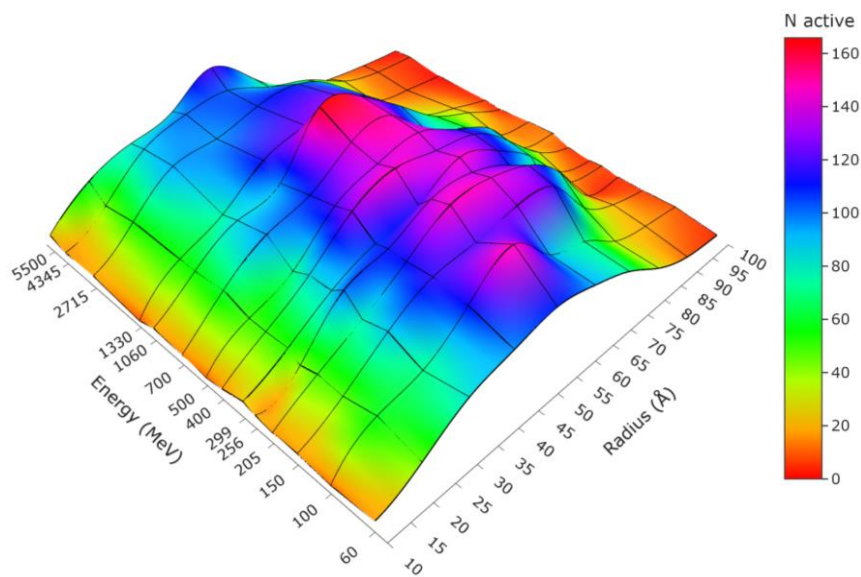

**Figure 8.** The surface of the radial distribution of chemically active carbon atoms in chains for different energies of the uranium ion.

## Conclusion

We simulated the kinetics of damage formation induced by swift heavy ions (SHI) decelerated in amorphous polyethylene in the electronic stopping regime. We coupled the MC TREKIS-3 code, which describes the excitation of the material around the ion trajectory, with the MD simulations (LAMMPS), which model the response of the atomic system to the excitation. This enabled us to obtain comprehensive insights into both the structural and chemical properties of the damaged polyethylene along the whole SHI trajectories (Xe and U ions).

Our results reveal a significant spatial separation of at least 10 micrometers between the positions of the maximum energy loss and maximum damage on the ion trajectory. We found a loop-like relationship between the produced damage intensity and the ion energy loss, and showed that the energy loss thresholds differ for ions realizing the opposite shoulders of the Bragg peak of the electronic energy loss of an SHI. Our study also showed that the position of the maximum of chemically active fragments on the ion path does not correspond to that of either the Bragg peak or the maximum of damaged carbons.

## Acknowledgements

The authors are grateful to Michael V. Sorokin and Sergey A. Gorbunov for helpful discussions. This work has been carried out using computing resources of the federal collective usage center Complex for Simulation and Data Processing for Mega-Science Facilities at NRC "Kurchatov Institute", http://ckp.nrcki.ru/.

## Author contributions

**P. Babaev:** Formal analysis, Methodology, Validation, Investigation, Writing – original draft, Visualization. **R. Voronkov:** Formal analysis, Writing – review & editing, Visualization. **A.E. Volkov:** Supervision, Conceptualization, Formal analysis, Writing – original draft, review & editing.

## Conflicts of interest or competing interests

The authors declare no conflict of interest, financial or otherwise.

## Data and code availability

Data will be made available on request.

## Supplementary information

Not Applicable.

# Ethical approval

Not Applicable.